\documentclass[12pt]{article}

\usepackage[dvips]{graphicx}
\usepackage{epsfig}
\usepackage{amsmath,amsfonts,amssymb,amsthm}
\usepackage{verbatim}
\usepackage{psfrag}
\usepackage{bm}
\usepackage{bbm}
\usepackage[square,comma,sort&compress,numbers]{natbib}
\usepackage{color}
\usepackage{slashed}
\usepackage{axodraw}

\usepackage{epsf,epsfig}
\usepackage{graphics}

\begin{document}
\thispagestyle{empty}

\begin{flushright}
{
\small
TUM-HEP-858-12\\
TTK-12-40
}
\end{flushright}

\vspace{0.4cm}
\begin{center}
\Large\bf\boldmath
Baryogenesis from Mixing of Lepton Doublets
\unboldmath
\end{center}

\vspace{0.4cm}

\begin{center}
{Bj\"orn~Garbrecht}\\
\vskip0.2cm
{\it Physik Department T70, James-Franck-Stra{\ss}e,\\
Technische Universit\"at M\"unchen, 85748 Garching, Germany\\
\vspace{0.15cm}
and Institut f\"ur Theoretische Teilchenphysik und Kosmologie,\\ 
RWTH Aachen University, 52056 Aachen, Germany}\\
\vskip1.4cm
\end{center}

\begin{abstract}
It is shown that the mixing of lepton doublets of the Standard
Model can yield sizeable contributions to the lepton asymmetry, that
is generated through the decays of right-handed neutrinos at
finite temperature in the early Universe. When calculating
the flavour-mixing correlations, we account for the effects of
Yukawa as well as of gauge interactions. We compare the
freeze-out asymmetry
from lepton-doublet mixing to the standard contributions
from the mixing and direct decays of right-handed neutrinos.
The asymmetry from lepton mixing is considerably large when the
mass ratio between the right-handed neutrinos is of order of a few,
while it becomes Maxwell-suppressed for larger hierarchies. For an
intermediate range between the case of
degenerate right-handed neutrinos
(resonant Leptogenesis) and the hierarchical case, lepton mixing
can yield the main contribution to the lepton asymmetry.
\end{abstract}



\section{Introduction}

Sources for $CP$-violating effects are often categorised into contributions
from mixing and from direct decays. This applies to Leptogenesis~\cite{Fukugita:1986hr} as well,
where usually the mixing~\cite{Covi:1996wh} and the direct
decays of right-handed singlet neutrinos
$N_i$ are accounted for. Of particular interest is the source from
mixing, because in the situation where the mass-difference of the
right-handed neutrinos is small, it gives rise to a resonantly enhanced
$CP$-asymmetry~\cite{Covi:1996wh,Flanz:1996fb,Pilaftsis:1997dr,Pilaftsis:1997jf,Pilaftsis:2003gt,Pilaftsis:2005rv,Garbrecht:2011aw,Garny:2011hg,Drewes:2012ma}.

Besides the right-handed neutrinos, Higgs bosons and Standard Model lepton
doublets directly take part in Leptogenesis. Direct
decay asymmetries from Higgs
bosons and lepton doublets are usually not considered, because in the
vacuum, these particles cannot decay into heavy right-handed neutrinos.
Consequently, at zero temperature, there are no kinematic cuts through
the propagators of a
right-handed neutrino and a lepton- or Higgs-doublet that can lead to
$CP$-violation. At finite temperature however, these cuts
contribute to the
asymmetry~\cite{Covi:1997dr,Giudice:2003jh,Garbrecht:2010sz},
since the plasma continuously absorbs and produces leptons and Higgs bosons.

Concerning mixing, right-handed neutrinos, as these are sterile, are well
suited for the efficient
production of asymmetries since flavour off-diagonal
(where we define the flavour of these particles by their mass-eigenstates,
in contrast to Standard Model leptons) 
correlations are only damped by the Yukawa couplings, which are small by
the requirement of a substantial deviation of the right-handed neutrinos from
equilibrium.

Gauge interactions, in particular in the Standard Model, are typically
much larger than the Yukawa couplings of the right-handed neutrinos. However,
as these are flavour blind, they do not directly damp off-diagonal flavour
correlations. Therefore, also the gauged particles within the Standard Model or its extensions can lead to substantial $CP$-violation from mixing.
For multiple Higgs doublets, it is shown in
Ref.~\cite{Garbrecht:2012qv}, that their mixing
can be a viable source for Leptogenesis. In the present work, we show that
the mixing of the lepton doublets of the Standard Model contributes
a sizeable amount of $CP$-violation to Leptogenesis as well.

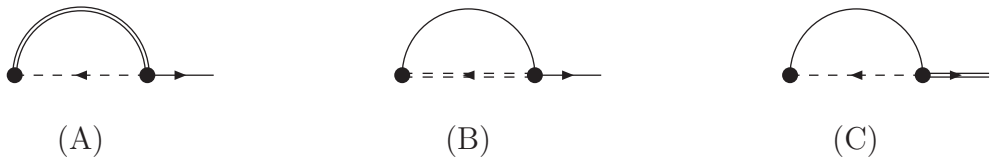
\begin{figure}[t!]
\begin{center}
\begin{picture}(100,60)(0,0)
\SetOffset(-25,0)
\CArc(50,25)(24.3,0,180)
\CArc(50,25)(25.7,0,180)
\DashArrowLine(75,25)(25,25){4}
\Vertex(25,25){3}
\Vertex(75,25){3}
\ArrowLine(75,25)(100,25)
\Text(50,0)[]{(A)}
\end{picture}
\hskip 1.5cm
\begin{picture}(100,60)(0,0)
\SetOffset(-25,0)
\CArc(50,25)(25,0,180)
\DashLine(75,24.3)(25,24.3){4}
\ArrowLine(51,25)(49,25)
\DashLine(75,25.7)(25,25.7){4}
\Vertex(25,25){3}
\Vertex(75,25){3}
\ArrowLine(75,25)(100,25)
\Text(50,0)[]{(B)}
\end{picture}
\hskip 1.5cm
\begin{picture}(100,60)(0,0)
\SetOffset(-25,0)
\CArc(50,25)(25,0,180)
\DashArrowLine(75,25)(25,25){4}
\Vertex(25,25){3}
\Vertex(75,25){3}
\Line(75,24.3)(100,24.3)
\ArrowLine(87,25)(88,25)
\Line(75,25.7)(100,25.7)
\Text(50,0)[]{(C)}
\end{picture}
\end{center}
\caption{\label{figure:collison:source}
Diagrammatic representation of contributions to the collision term that yield a
source for the lepton asymmetry (A) through mixing of right-handed neutrinos, (B) through
mixing of Higgs doublets and (C) through mixing of lepton doublets. Solid
lines with arrows are propagators for the Standard Model lepton
doublets $\ell$, dashed lines with arrows
are propagators for Higgs doublets $\phi$
and solid lines without arrows are propagators for the right-handed
neutrinos $N$. The double lines represent resummed propagators
that in particular account for the mixing of flavour.}
\end{figure}

In order to calculate the generation of the asymmetries in the
finite temperature background, we use the Closed-Time-Path (CTP)
method~\cite{Schwinger:1960qe,Keldysh:1964ud,Calzetta:1986cq,Prokopec:2003pj,Prokopec:2004ic,Garbrecht:2010sz,Garbrecht:2012qv,Garbrecht:2011aw,Garny:2011hg,Drewes:2012ma,Buchmuller:2000nd,De Simone:2007rw,Garny:2009rv,Garny:2009qn,Anisimov:2010aq,Garny:2010nj,Beneke:2010wd,Beneke:2010dz,Garny:2010nz,Anisimov:2010dk,Cirigliano:2009yt,Cirigliano:2011di}.
The diagrammatic representation of the source terms for the asymmetry in
Figure~\ref{figure:collison:source} indicates how the mixing of
the various species yields contributions to
Leptogenesis . When we assume that the
right-handed neutrinos are heavier than the lepton doublets and Higgs bosons,
the imaginary parts of the self-energies that appear within the resummed
propagators in Figure~\ref{figure:collison:source}(B)
and~\ref{figure:collison:source}(C) are purely thermal,
as the corresponding
$1\leftrightarrow 2$ processes are kinematically forbidden in the vacuum.
This resembles the situation for Leptogenesis
from the mixing of right-handed neutrinos with masses of the
${\rm GeV}$ scale or
below~\cite{Drewes:2012ma,Akhmedov:1998qx,Asaka:2005pn,Canetti:2012vf,Canetti:2012kh},
where the $CP$-violating cut
is dominated by purely thermal contributions~\cite{Drewes:2012ma}.
For the direct $CP$-violation, the contribution of such non-standard
cuts is calculated in Ref.~\cite{Garbrecht:2010sz}.

In Section~\ref{sec:off:corr}, we derive the off-diagonal correlations
within the distributions of the left-handed lepton doublets. Even though
gauge interactions do not directly violate flavour and damp its correlations,
they catalyse the decay of flavour correlations in the presence
of flavour-dependent masses or interactions. For flavoured
Leptogenesis, the relevance of such effects is discussed in
Ref.~\cite{Beneke:2010dz}. Here, we show that gauge interactions effectively
limit the maximal resonant enhancement of a $CP$-asymmetry resulting from
the mixing of gauged particles. For the mixing of Higgs doublets,
this aspect is not discussed in Ref.~\cite{Garbrecht:2012qv}, a
shortcoming that is addressed in Section~\ref{sec:Higgs:mix}.

An important aspect within the present context is the role of
lepton-number violation and conservation. The spinor trace of
diagram~\ref{figure:collison:source}(c), that quantifies the
rate of $CP$-violation from mixing of lepton doublets, contains no
lepton-number violation that could be described by insertions of
an odd number of
Majorana mass-terms. Therefore, lepton number must in some sense be
conserved. This, we show in Section~\ref{sec:cancellation}, where
it turns out that the source for the asymmetry in the lepton doublets
cancels the helicity asymmetry within the right-handed neutrinos.
The latter is however rapidly violated by the Majorana masses, provided
the right-handed neutrinos are non-relativistic, as it is the case
in the strong washout regime.

In Section~\ref{sec:example:strongwashout}, we then calculate the
freeze-out value of the asymmetry that results from mixing lepton
doublets in the strong washout regime. This, we compare to the asymmetry
from the standard sources, {\it i.e.} from the
mixing and direct decays of right-handed
neutrinos. For illustrative purposes, we choose a particular slice
in parameter space and vary the mass ratio of the right-handed
neutrinos between one (degenerate case) and a few. It is a characteristic
feature of the asymmetry from lepton mixing, that it is exponentially
small for large mass ratios, as the $CP$-violating cut is purely
thermal and becomes Maxwell suppressed in that situation.

We close with summarising and concluding remarks, that are given in
Section~\ref{sec:sum:disc}.

\section{Off-Diagonal Correlations of Lepton Doublets}
\label{sec:off:corr}

The model that we consider here is the same that underlies flavoured Leptogenesis~\cite{Beneke:2010dz,Endoh:2003mz,Abada:2006fw,Nardi:2006fx} and is given by
the Lagrangian
\begin{align}
\label{Lagrangian}
{\cal L}=&\frac{1}{2}\bar\psi_{Ni}({\rm i} \partial\!\!\!/-M_{Nij}) \psi_{Nj}
+\bar\psi_{\ell a}{\rm i}\partial\!\!\!/\psi_{\ell a}
+(\partial^\mu\phi^\dagger)(\partial_\mu \phi)
\\\notag
-&Y_{ia}^*\bar\psi_{\ell a} \tilde\phi P_{\rm R}\psi_{Ni}
-Y_{ia}\bar\psi_{Ni}P_{\rm L}\tilde\phi^\dagger\psi_{\ell a}
-h_{ab}\phi^\dagger \bar\psi_{{\rm R}a} P_{\rm L} \psi_{\ell b}
-h_{ab}^*\phi \bar\psi_{\ell b} P_{\rm R} \psi_{{\rm R} a}
\,.
\end{align}
The Standard Model Higgs doublet is denoted by $\phi$, the lepton doublets
by $\ell_a$, right-handed charged
leptons by ${\rm R}_a$ and right-handed neutrinos
by $N_i$. The associated spinors are denoted by $\psi$ with the subscript
of the respective field. We use the definition
$\tilde\phi_a=\phi_b^*\epsilon_{ba}$, but suppress explicit
indices of the ${\rm SU}(2)_{\rm L}$ gauge group in the
following.
We transform the $N_i$ to a basis where $M_N$ is diagonal and write
$M_{Ni}=M_{Nii}$.
Furthermore,
we take $h_{ab}$ to be diagonal, where $a,b=e,\mu,\tau$.
We are mainly interested in the dynamics within the temperature
range $10^9\,{\rm GeV}<T<10^{12}\,{\rm GeV}$, such that only the $\tau$-lepton
Yukawa couplings are in equilibrium. In that case, we can neglect $h_{ee}$
and $h_{\mu\mu}$ and then perform a unitary transformation of the lepton flavours $\ell_e$ and $\ell_\mu$ to $\ell_{\sigma\perp}$
and $\ell_\sigma$
such that $Y_{2\sigma_\perp}=Y_{3\sigma_\perp}=0$, see
Ref.~\cite{Beneke:2010dz} for the explicit construction of such a transformation.
(After this transformation, $h$ is no longer diagonal within the block
of the flavours $e$ and $\tau$. As we may however neglect these Yukawa
couplings at temperatures above $10^9\,{\rm GeV}$, $h$ can still
be considered as effectively diagonal).
Then,
only lepton flavour asymmetries within $\ell_\tau$ and $\ell_\sigma$ are produced, while there is no asymmetry in the remaining
linear combination $\ell_{\sigma\perp}$, since $\ell_{\sigma\perp}$ couples only
to one of the right-handed neutrinos,
and no $CP$-violating interference terms occur. While
it is the salient feature of flavoured Leptogenesis that
all $CP$-odd off-diagonal
correlations between the
flavours $\tau$ and $\sigma$ are rapidly erased, because the interactions mediated by  $h_{\tau\tau}$ are
in equilibrium, we find in the present work that the deviation of
the right-handed neutrinos from equilibrium does generically sustain
$CP$-even [{\it cf.} Eq.~(\ref{lepcon:orderYsqr}) below]
off-diagonal correlations of the
leptons [Eq.~(\ref{sol:deltan}) below], which in turn
enter the source term for the $CP$-violating lepton asymmetry.
We also note that
the present discussion does not directly apply but can easily be
generalised to temperatures below $10^9\,{\rm GeV}$, where the
flavours $e$ and $\mu$ are distinguishable.

The present calculations based on the CTP formalism require numerous definitions
that can be found within Refs.~\cite{Beneke:2010wd,Beneke:2010dz,Garbrecht:2011aw}.
In particular ${\rm i}S^{<,>}$ denotes fermionic and ${\rm i}\Delta^{<,>}$
scalar Wightman functions. Subscripts on these indicate the particular fields
and flavour correlations. Distribution functions for particles and
anti-particles are denoted by $f$ and $\bar f$, number densities by $n$
and $\bar n$. Again,
fields and flavour correlations are indicated by subscripts. A prefix $\delta$
implies that the particular quantity is the deviation from the equilibrium
propagator, distribution function or density.

The dynamics of the out-of equilibrium distributions of left-handed
lepton doublets can be described by the
kinetic equation~\cite{Beneke:2010dz}
\begin{align}
\label{kin:eq}
{\rm i}\partial_\eta g^{<,>}_{hab}
+h \left(\varsigma_{aa}^{\slashed{\rm fl}}-\varsigma_{bb}^{\slashed{\rm fl}}\right)
g^{<,>}_{hab}=-\frac14 {\rm tr}
\left[
{\rm i}{\cal C}_\ell+{\rm i}{\cal C}_\ell^\dagger
\right]_{ab}
\,,
\end{align}
where $\eta$ is the conformal time in the expanding Universe.
The functions $g^{<,>}_h$ appear within the spinor components of the
left-handed lepton propagator $S^{<,>}_\ell$ and factorise into
statistical and spectral terms in the narrow width
limit~\cite{Kainulainen:2001cn,Prokopec:2003pj,Prokopec:2004ic,Garbrecht:2002pd,Garbrecht:2011aw}.
When $|\mathbf k|\gg \sqrt G T$, where
$G=(3/2)g_2^2+(1/2)g_1^2$, we may approximate
\begin{subequations}
\label{gell:onshell}
\begin{align}
g_{-ab}^{<,>}(k)=&-2\pi\delta(k^2)|\mathbf k|\vartheta(k^0)\delta f_{\ell ab}(\mathbf k)\,,\\
g_{+ab}^{<,>}(k)=&2\pi\delta(k^2)|\mathbf k|\vartheta(-k^0)\delta \bar f_{\ell ab}(\mathbf k)\,.
\end{align}
\end{subequations}
For $|\mathbf k|\ll \sqrt G T$, this approximation, in particular the 
relation between the sign of $k^0$ and the helicity, becomes inaccurate due
to the presence of hole modes. However, this region of soft
momenta covers only a small part of the phase-space,
such that the
approximations~(\ref{gell:onshell})
are sufficiently accurate for the present purposes.
The flavour-dependent self-mass terms $\varsigma^{\slashed{\rm fl}}$ are
given by
\begin{align}
\varsigma_{ab}^{\slashed{\rm fl}}(k)
=\frac{[h^\dagger h]_{ab} T^2}{16|\mathbf k|}
\,,
\end{align}
where we have neglected the contributions from the couplings $Y$.

It is useful to decompose the collision term as
\begin{align}
{\cal C}_\ell=
{\cal C}_\ell^{CP \rm V}
+{\cal C}_\ell^{Y}
+{\cal C}_\ell^{h}
+{\cal C}_\ell^{g}
\end{align}
into contributions mediated by the Yukawa couplings $Y$ and $h$
and the gauge couplings $g$.
Within ${\cal C}_\ell^Y$, we collect the $CP$-conserving
processes mediated by $Y$, which are to leading
approximation quadratic in $Y$.
The term ${\cal C}_\ell^{CP \rm V}$ is mediated
by $Y$ as well and accounts for $CP$-violating processes.
To leading approximation,
it is fourth order in $Y$.

The part of the lepton self-energy mediated by $Y$ is
\begin{align}
{\rm i}\slashed\Sigma_{\ell ab}^{Y fg}(k)
=Y_{ai}^\dagger Y_{jb}
\int\frac{d^4 p}{(2\pi)^4}P_{\rm R}{\rm i}S^{fg}_{N ij}(p)
{\rm i}\Delta_\phi^{gf}(p-k)\,,
\end{align}
where $f,g$ are CTP indices.
The term ${\cal C}_\ell^{Y}$ is of importance for two reasons:
First, as in the standard scenario of Leptogenesis, its
diagonal components describe
the washout of the $CP$-odd lepton asymmetries accumulated in the
diagonal correlations of lepton flavours. (For flavoured
Leptogenesis, the off-diagonal
correlations are rapidly erased due to the Yukawa coupling $h_{\tau\tau}$.)
Second, its off-diagonal components correspond to the source
of $CP$-even, flavour off-diagonal correlations in the
lepton doublets. As the $CP$-odd off-diagonal correlations are
suppressed by the Yukawa interactions $h_{\tau\tau}$, and the
$CP$-even off-diagonal correlations are suppressed by an order
of $Y^2$ times the deviation of the right-handed neutrinos
from equilibrium, when compared to the diagonal distribution
functions, we can substitute the flavour diagonal components
of ${\rm i}S^{<,>}_{\ell}$ for the purpose of calculating the
contributions to ${\cal C}_\ell^{Y}$ that are of leading importance.
We thus obtain
\begin{align}
\label{coll:Y}
{\cal C}^{Y}_{\ell ab}(k)+&{\cal C}_{\ell ab}^{CP \rm V}
={\rm i}\slashed\Sigma_{\ell ac}^{Y>}(k){\rm i}S^<_{\ell cb}(k)
-{\rm i}\slashed\Sigma_{\ell ac}^{Y<}(k){\rm i}S^>_{\ell cb}(k)
\\\notag
\approx&
-[Y^\dagger Y]_{ab}
P_{\rm R}\int\frac{d^4 p}{(2\pi)^4}
\left[
{\rm i}S^>_{Nii}(p)
{\rm i}\Delta^<_\phi(p-k){\rm i}S^<_{\ell bb}(k)
-
{\rm i}S^<_{Nii}(p)
{\rm i}\Delta^>_\phi(p-k){\rm i}S^>_{\ell bb}(k)
\right]
\\\notag
+&{\cal C}_{\ell ab}^{CP \rm V}
\,,
\end{align}
where we take ${\rm i}\Delta^{<,>}_\phi$
to be of equilibrium form. Note that
while the off-diagonal correlations within $\ell$ and $N$,
which are always out-of-equilibrium, can be neglected for calculating
${\cal C}_\ell^Y$ to second order in $Y$, within the $CP$ violating
source ${\cal C}_\ell^{CP \rm V}$, these off-diagonal correlations
are essential
for the asymmetry that we calculate in Eq.~(\ref{source:lepto})
below.

For definiteness, we assume strong washout, {\it i.e.} $M_{Ni}\gg T$
at the time of Leptogenesis. (Due to the
exponential Maxwell-suppression, masses of a factor of a few above the
temperature are enough for sufficiently accurate approximations.)
This allows us to neglect quantum statistical
blocking and enhancement factors on many occasions and to use
Maxwell distributions instead of Fermi-Dirac or Bose-Einstein
distributions.
We parametrise the deviation of the neutrino distribution
from equilibrium by a pseudo-chemical potential
$\mu_{Ni}$ as
\begin{align}
\delta f_{Ni}(p)=\frac{\mu_{Ni}}{T}{\rm e}^{-\beta\sqrt{\mathbf p^2+M_{Ni}^2}}\,.
\end{align}
While in the present setup, there are no interactions that directly force
kinetic equilibrium of the sterile neutrinos, it turns out that
above parametrisation indeed gives a good approximation to the
distribution function of sterile neutrinos in strong washout
scenarios~\cite{Beneke:2010wd,Basboll:2006yx}.

We now aim to compute the off-diagonal distributions $f_{\ell ab}$ and
$\bar f_{\ell ab}$, $a\not=b$. In kinetic equilibrium, these can be inferred from
the number densities of leptons and anti-leptons,
\begin{subequations}
\begin{align}
\delta n^+_{\ell ab}&=-2\int\limits_0^\infty\frac{dk^0}{2\pi}
\int\frac{d^3k}{(2\pi)^3}g^{<,>}_{-ab}(k)\,,
\\
\delta n^-_{\ell ab}&=-2\int\limits_{-\infty}^0\frac{dk^0}{2\pi}
\int\frac{d^3k}{(2\pi)^3}g^{<,>}_{+ab}(k)\,.
\end{align}
\end{subequations}
(Note that $\pm$ on $\delta n^{\pm}_{\ell ab}$ refers to the 
particle and anti-particle, whereas on $g^{<,>}_{\pm ab}$, it refers
to helicity.)
The prefix $\delta$ indicates that these quantities are deviations of the
number densities from their equilibrium values. Consequently, the
charge density $q_{\ell ab}=\delta n^+_{\ell ab}-\delta n^-_{\ell ab}$
can deviate from zero.
Pair creation and annihilation processes tend to strongly suppress
the combination $\delta n^+_{\ell ab}+\delta n^-_{\ell ab}$. However,
it turns out that for small Yukawa couplings, the finite value of this
combination limits the maximal resonant enhancement of the asymmetry,
{\it cf.} Eq.~(\ref{sol:deltan}) below.
For this reason, equations for $\delta n^{\pm}_{\ell ab}$ rather than
just $q_{\ell ab}$ need to be derived, as a generalisation of the
methods introduced in Ref.~\cite{Garbrecht:2012qv}.

In order to obtain these densities
when the time-derivative in the kinetic equation~(\ref{kin:eq})
is small compared to the effective mass difference
$\zeta^{\slashed{\rm fl}}_{aa}-\zeta^{\slashed{\rm fl}}_{bb}$, it is useful to multiply these
equations by $|\mathbf k|$ and then to integrate over $d^4 k$.
Doing so, we encounter on the right-hand side
\begin{align}
-\int\limits_0^\infty\frac{dk^0}{2\pi}\int\frac{d^3k}{(2\pi)^3}|\mathbf k|{\rm tr}{\cal C}^{Y}_{\ell}(k)
=
\int\limits_{-\infty}^0\frac{dk^0}{2\pi}\int\frac{d^3k}{(2\pi)^3}|\mathbf k|{\rm tr}{\cal C}^{Y}_{\ell}(k)
=:-\sum\limits_i Y^\dagger_{ai} Y_{ib}B^Y_i\,.
\end{align}
In the limit of strong washout, where the quantum statistical distributions
can be approximated by Maxwell distributions, we obtain
\begin{align}
\label{B:Y}
B^Y_i\approx
-\frac{T^{\frac32}M_{Ni}^\frac72}{2^\frac{13}{2}\pi^\frac52}
\frac{\mu_{Ni}}{T}{\rm e}^{-\frac{M_{Ni}}{T}}\,.
\end{align}
Clearly, this term is exponentially suppressed for large $M_{Ni}$.
It can be interpreted as the $CP$-violating cut that appears in the resummed
lepton propagator of Figure~\ref{figure:collison:source}(C), and since
the reaction $\ell\leftrightarrow\phi^*+N_i$ is kinematically
forbidden in the vacuum, this cut is purely thermal.

The interactions mediated by the Yukawa couplings $h$ lead to the
decay of off-diagonal correlations of the lepton doublets.
The relevant contribution to the collision term is given by~\cite{Beneke:2010dz}
\begin{align}
{\cal C}^{h}_{\ell ab}(k)
\approx&
h^\dagger_{ac}h_{cd}
\int\frac{d^4p}{(2\pi)^4}
\left[
{\rm i}S^>_{\rm R cc}(p){\rm i}\Delta^>_\phi(k-p)
-{\rm i}S^<_{\rm R cc}(p){\rm i}\Delta^<_\phi(k-p)
\right]{\rm i}\delta S_{\ell cb}(k)\,.
\end{align}
In the integrated kinetic equations, then the following terms
occur:
\begin{align}
-\int\limits_0^\infty
\frac{dk^0}{2\pi}&
\int\frac{d^3k}{(2\pi)^3}|\mathbf k|{\rm tr}{\cal C}^{h}_{\ell ab}(k)
=
-h^\dagger_{ac}h_{cd}
\int\limits_0^\infty
\frac{dk^0}{2\pi}
\int\frac{d^3k}{(2\pi)^3}
\int\frac{d^4p}{(2\pi)^4}
|\mathbf k|
\\\notag\times&
{\rm tr}\left[
{\rm i}S^>_{\rm R cc}(p){\rm i}\Delta^>_\phi(k-p)
-{\rm i}S^<_{\rm R cc}(p){\rm i}\Delta^<_\phi(k-p)
\right]{\rm i}\delta S_{\ell cb}(k)
\\\notag
=&-h^\dagger_{ac}h_{cd} \delta n^+_{\ell ab}
\int\limits_0^\infty
\frac{dk^0}{2\pi}
\int\frac{d^3k}{(2\pi)^3}
\int\frac{d^4p}{(2\pi)^4}
|\mathbf k|
\\\notag\times&
{\rm tr}\left[
{\rm i}S^>_{\rm R cc}(p){\rm i}\Delta^>_\phi(k-p)
-{\rm i}S^<_{\rm R cc}(p){\rm i}\Delta^<_\phi(k-p)
\right]
2 S^{\cal A}_{\ell cb}(k)
\frac{12\beta^3{\rm e}^{\beta k^0}}{({\rm e}^{\beta k^0}+1)^2}
\\\notag
&=:-B_\ell^{\slashed{\rm fl}} \left[h^\dagger h \delta n^+_\ell\right]_{ab}
\end{align}
and likewise
\begin{align}
\int\limits_{-\infty}^0
\frac{dk^0}{2\pi}
\int\frac{d^3k}{(2\pi)^3}|\mathbf k|{\rm tr}{\cal C}^{h}_{\ell ab}(k)
&=-B_\ell^{\slashed{\rm fl}} \left[h^\dagger h \delta n^-_\ell\right]_{ab}\,.
\end{align}
With these coefficients, the integrated kinetic equations are
\begin{subequations}
\label{eff:kin}
\begin{align}
\label{eff:kin:plus}
\frac{54\zeta(3) T}{\pi^2}\partial_\eta\delta n_{\ell ab}^+
&+{\rm i}\frac{\left(h_{aa}^2-h_{bb}^2\right)T^2}{16}\delta n^+_{\ell ab}
\\\notag
&=
-\sum\limits_i Y^\dagger_{ai}Y_{ib} B_i^Y
-(h_{aa}^2+h_{bb}^2)B_\ell^{\slashed{\rm fl}}\delta n^+_{\ell ab}
-B_\ell^g (\delta n^+_{\ell ab}+\delta n^-_{\ell ab})\,,
\\
\label{eff:kin:minus}
\frac{54\zeta(3) T}{\pi^2}\partial_\eta\delta n_{\ell ab}^-
&-{\rm i}\frac{\left(h_{aa}^2-h_{bb}^2\right)T^2}{16}\delta n^-_{\ell ab}
\\\notag
&=
-\sum\limits_i Y^\dagger_{ai}Y_{ib} B_i^Y
-(h_{aa}^2+h_{bb}^2)B_\ell^{\slashed{\rm fl}}\delta n^-_{\ell ab}
-B_\ell^g (\delta n^+_{\ell ab}+\delta n^-_{\ell ab})\,.
\end{align}
\end{subequations}
In principle, there also occur terms
\begin{align}
\propto([Y^\dagger Y]_{aa}+[Y^\dagger Y]_{bb})\delta n^\pm_{\ell ab}
\end{align}
on the right hand sides of these equations. However, we assume here that the dominant
flavour-sensitive interaction is mediated by the lepton-Yukawa couplings.
In particular, $h_{\tau\tau}$ dominates over the relevant elements of $Y$,
which is valid provided Leptogenesis occurs at temperatures below
$10^{12}\,{\rm GeV}$.

The coefficient $B_\ell^{\slashed{\rm fl}}$ is the averaged rate for
flavour-sensitive
processes that are mediated by the coupling $h$. As the Higgs boson
as well as the Standard Model leptons are massless in the symmetric
Electroweak phase, these reactions require the radiation of an additional 
gauge boson. A systematic calculation is currently on the way, but
from the similar problem of the production of light right-handed
neutrinos~\cite{Anisimov:2010gy,Salvio:2011sf,Laine:2011xm,Laine:2011pq,Besak:2012qm},
we may estimate that $B_\ell^{\slashed{\rm fl}}=1.0\times10^{-2}T^2$,
see the Appendix~\ref{app:rates}.

Pair creation and annihilation processes may change $\delta n_{\ell ab}^{\pm}$.
Even though these processes are flavour blind, we will see below that in
interplay with flavour-sensitive processes, they can be important for
the suppression of flavour off-diagonal correlations. Accounting
for the large number of Standard Model degrees of freedom, these processes
are dominated by the $s$-channel exchange of gauge bosons, and we estimate these
within Appendix~\ref{app:rates} as $B_\ell^{\rm g}=1.7\times10^{-3}T^2$.

When neglecting the time-derivative, Eqs.~(\ref{eff:kin}) yield the
induced off-diagonal ($a\not=b$) correlations
\begin{subequations}
\label{sol:deltan}
\begin{align}
q_{\ell ab}\equiv q_{\ell ab}(\eta\to\infty)&
\\\notag
=\delta n_{\ell ab}^+-\delta n_{\ell ab}^-
&=
{\rm i}
\frac
{(h_{aa}^2-h_{bb}^2)(T^2/8)\sum_i Y_{ai}^\dagger Y_{ib} B_i^Y}
{\left[(h_{aa}^2-h_{bb}^2)T^2/16\right]^2+(h_{aa}^2+h_{bb}^2)B_\ell^{\slashed{\rm fl}}[2 B_\ell^g+(h_{aa}^2+h_{bb}^2)B_\ell^{\slashed{\rm fl}}]}
\\\notag
&=:({\cal Q}_{\ell ab}/T^2)\sum_i Y_{ai}^\dagger Y_{ib} B_i^Y
\,,
\\
\delta n_{\ell ab}^+ +\delta n_{\ell ab}^-
&=
-
\frac
{2(h_{aa}^2+h_{bb}^2)B_\ell^{\slashed{\rm fl}} \sum_i Y_{ai}^\dagger Y_{ib} B_i^Y}
{\left[(h_{aa}^2-h_{bb}^2)T^2/16\right]^2+(h_{aa}^2+h_{bb}^2)B_\ell^{\slashed{\rm fl}}[2 B_\ell^g+(h_{aa}^2+h_{bb}^2)B_\ell^{\slashed{\rm fl}}]}
\,.
\end{align}
\end{subequations}
The general solution to Eqs.~(\ref{eff:kin}) involves the damping rates
\begin{align}
\Gamma^{\pm}_{q_{\ell ab}}
=\frac{\pi^2}{54\zeta(3) T}\left(
B_\ell^g
+(h_{aa}^2+h_{bb}^2)B_\ell^{\slashed{\rm fl}}
\pm\sqrt{{B_\ell^g}^2-\left[(h_{aa}^2-h_{bb}^2)T^2/16\right]^2}
\right)\,,
\end{align}
and is given by
\begin{align}
q_{\ell ab}(\eta)=q_{\ell ab}(\eta\to\infty)+
Q_1 {\rm e}^{-\Gamma^+_{q_{\ell ab}}\eta}
+Q_2 {\rm e}^{-\Gamma^-_{q_{\ell ab}}\eta}\,.
\end{align}
This implies that it takes a finite time for these
$CP$-even off-diagonal correlations to build 
up. In the present case, the rate for this is given by
$\Gamma^{-}_{q_{\ell \tau\tau_\perp}}\sim h_{\tau\tau}^2 T$.
This in particular implies that in the regime of flavoured Leptogenesis
below $10^{12}\,{\rm GeV}$, when the interactions mediated by $h_{\tau\tau}$
are in equilibrium, flavour off-diagonal lepton correlations are present,
and their magnitude is given by Eqs.~(\ref{sol:deltan}).

We express the resonant enhancement of the asymmetries
through the factor
\begin{align}
\label{Qlab}
{\cal Q}_{\ell ab}=\frac
{(h_{aa}^2-h_{bb}^2)(T^4/8)}
{\left[(h_{aa}^2-h_{bb}^2)/16\right]^2 T^4+(h_{aa}^2+h_{bb}^2)B^{\slashed{\rm fl}}[2 B^g+(h_{aa}^2+h_{bb}^2)B^{\slashed{\rm fl}}]}
\,.
\end{align}
The fact that $\delta n_{\ell ab}^+ +\delta n_{\ell ab}^-$ is
suppressed but
not vanishing implies that kinetic equilibrium mediated by pair
creation and annihilation processes is only approximately established.
Processes of kinetic equilibration that do not mediate
between $\delta n_\ell^+$ and $\delta n_\ell^-$, {\it i.e.}
which do not correspond to pair creation and annihilation,
are however not conflicting with the solution~(\ref{sol:deltan}).
We therefore describe the distribution functions in terms of
the number densities as
\begin{subequations}
\label{lepdist:kineq}
\begin{align}
\delta f_\ell(\mathbf k)=&\frac{12\delta n_\ell^+}{T^3}\frac{{\rm e}^{|\mathbf k|/T}}{\left({\rm e}^{|\mathbf k|/T}+1\right)^2}\,,
\\
\delta \bar f_\ell(\mathbf k)=&\frac{12\delta n_\ell^-}{T^3}\frac{{\rm e}^{|\mathbf k|/T}}{\left({\rm e}^{|\mathbf k|/T}+1\right)^2}
\,.
\end{align}
\end{subequations}
Notice also that from Eqs.~(\ref{sol:deltan}) it follows that
$n^+_{\ell ab}=n^{-*}_{\ell ab}$,
$n^\pm_{\ell ab}=n^{\pm *}_{\ell ba}$ and therefore
$n^+_{\ell ab}=n^-_{\ell ba}$.
This implies that the flavour off-diagonal lepton correlations
are $CP$-even,
\begin{align}
\label{lepcon:orderYsqr}
P_{\rm L}{\rm i}S^{fg}_{\ell ab}(p)
=C\left[P_{\rm L}{\rm i}S^{gf}_{\ell ba}(-p)\right]^t C^\dagger +{\cal O}(Y^4)\,,
\end{align}
such that to this end, the lepton asymmetry is not yet broken.

\section{Asymmetries of Lepton Doublets and Singlet Neutrinos}
\label{sec:cancellation}

In this Section, we calculate the $CP$-odd charge asymmetries within
the diagonal components
of the distribution functions of lepton doublets as well as the
helicity asymmetries within the singlet neutrinos, that are induced
by the off-diagonal, $CP$-even correlations of lepton doublets.
In the strong washout scenario, helicity asymmetries within the
right-handed
neutrinos are quantitatively irrelevant, because of the Majorana masses, which
effectively erase such asymmetries when the right-handed
neutrinos decay. However, helicity
asymmetries can be of crucial importance in weak washout scenarios~\cite{Drewes:2012ma,Akhmedov:1998qx,Asaka:2005pn,Canetti:2012vf,Canetti:2012kh}.

In the present context, we are interested in the helicity asymmetries,
because the lepton-doublet propagator and its self-energy correction do not
encompass lepton-number violating combinations of
operators, such that the total source for the
lepton asymmetry plus the helicity asymmetry in the right-handed neutrinos
should vanish. This constraint provides a useful consistency check
that we perform in the present Section.
In spite of the vanishing of the total source, the
asymmetry of the singlet neutrinos is completely washed out in the decays, that
proceed via the Majorana mass term within the strong washout
regime. In contrast, the active leptons experience
finite washout rates from the inverse decays, such that a total lepton
asymmetry remains at freeze out.

The self-energy for the right-handed neutrino is~\cite{Beneke:2010wd,Garbrecht:2011aw,Drewes:2012ma}
\begin{align}
{\rm i}\slashed\Sigma^{fg}_{Nij}(k)
=&g_w Y_{ia}Y^\dagger_{bj}\int\frac{d^4p}{(2\pi)^4}
P_{\rm L}{\rm i}S^{fg}_{\ell ab}(p){\rm i}\Delta_\phi^{fg}(k-p)
\\\notag
+&g_w Y_{ia}^*Y^t_{bj}\int\frac{d^4p}{(2\pi)^4}
C\left[P_{\rm L}{\rm i}S^{gf}_{\ell ba}(-p)\right]^tC^\dagger{\rm i}\Delta_\phi^{gf}(p-k)
\,,
\end{align}
where $g_w=2$ is the dimension of the ${\rm SU}(2)_{\rm L}$ representation
of the lepton doublet. The self-energy
enters into the collision term for the right-handed neutrino as
\begin{align}
{\rm tr}{\cal C}_N(k)=&{\rm tr}\left[{\rm i}\slashed\Sigma_N^>(k){\rm i}S_N^<(k)
-{\rm i}\slashed\Sigma_N^<(k){\rm i}S_N^>(k)\right]\,.
\end{align}
A $CP$- and helicity-violating contribution ${\cal C}_N^{CP \rm V}$
is induced by the flavour off-diagonal correlations of the lepton
doublets. We denote this as the source term
\begin{align}
{\cal S}_{N ij}=
&\int\frac{d^4 k}{(2\pi)^4}\frac14{\rm tr}
\left[
{\cal C}^{CP \rm V}_N(k)
+{\cal C}^{{CP \rm V}\dagger}_N(k)
\right]_{ij}
\\\notag
=&
\frac12\sum\limits_{\underset{a\not=b}{a,b}}
\int\frac{d^4 k}{(2\pi)^4}
\int\frac{d^4 p}{(2\pi)^4}
g_w Y_{ia}Y^\dagger_{bj}
{\rm tr}
\Big[
{\rm i}\delta S_{\ell ab}(p)
\big(
{\rm i}\Delta_\phi^>(k-p)
\left[
{\rm i}S^<_{Nii}(k)+{\rm i}S^<_{Njj}(k)
\right]
\\\notag
&\hskip5.7cm
-{\rm i}\Delta_\phi^<(k-p)
\left[
{\rm i}S^>_{Nii}(k)+{\rm i}S^>_{Njj}(k)
\right]
\big)
\Big]\,.
\end{align}
We have made use here of the ${\cal O}(Y^2)$
lepton number conservation, Eq.~(\ref{lepcon:orderYsqr}), and
assumed that there is no charge asymmetry within the Higgs field,
such that ${\rm i}\Delta_\phi^{<,>}(p)={\rm i}\Delta_\phi^{>,<}(-p)$.
Moreover, the off-diagonal correlations within the right-handed
neutrinos, which are important for the
usually considered $CP$-violating sources
for Leptogenesis, are neglected here. (An example with a quantitative comparison of the source from mixing and decays
of right-handed neutrinos and the source from mixing lepton doublets
is presented in Section~\ref{sec:example:strongwashout}.)

The source term can be expressed to enter the integrated
kinetic equation
for the right-handed neutrinos in the form
\begin{align}
\frac 12
\partial_\eta\int\frac{d^4 k}{(2\pi)^4}{\rm i}\,{\rm tr}[\gamma^0 S^{<,>}_{N ij}(k)]
=-{\cal S}_{N ij}\,.
\end{align}
The integral on the left hand side corresponds to the rate
of producing a helicity asymmetry,
which is in the limit of vanishing Majorana masses the same as an asymmetry
between right-handed neutrinos and anti-neutrinos. Note that this
integrated kinetic equation does not account for the total production
of singlet neutrinos. In order to obtain these rates, the collision
term is to be integrated from $0$ to $\infty$ rather than
from $-\infty$ to $\infty$ over $dk^0$~\cite{Beneke:2010wd}.
The factor $\frac 12$ on the
right hand side is a suitable normalisation, because in writing the
Majorana field as a four component spinor, the number of degrees
of freedom is doubled, though they are related by the Majorana
constraint.

As for the lepton collision term ${\cal C}_\ell^Y$,
the approximation~(\ref{coll:Y}) is used in order
to calculate the $CP$-even
off-diagonal correlations in terms of the deviation of the
right-handed neutrinos from equilibrium. For that purpose, we have taken
$S_\ell$ to be of equilibrium form under the integral.
Now, we want to obtain the $CP$-odd asymmetry
in the lepton doublets that is induced by these off-diagonal correlations.
It follows from integrating ${\cal C}_\ell^{CP\rm V}$ as
\begin{align}
\label{source:lepto}
{\cal S}_{\ell ab}=&\int\frac{d^4 k}{(2\pi)^4}\frac12{\rm tr}
\left[
{\cal C}^{CP\rm V}_\ell(k)
+{\cal C}^{{CP\rm V}\dagger}_\ell(k)
\right]_{ab}
\\\notag
=&
\frac12\sum\limits_{\underset{c\not=b}{i,c}}
\int\frac{d^4 k}{(2\pi)^4}
\int\frac{d^4 p}{(2\pi)^4}
{\rm tr}
\Big[
\left(
P_{\rm R} {\rm i} S_{N ii}^>(p) {\rm i}\Delta_\phi^<(p-k)
-P_{\rm R} {\rm i} S_{N ii}^<(p) {\rm i}\Delta_\phi^>(p-k)
\right)
\\\notag&
\hskip4.3cm
\times
(Y^\dagger_{ai}Y_{ic}{\rm i}\delta S_{\ell cb}(k)
+{\rm i}\delta S_{\ell ac}(k)Y^\dagger_{ci}Y_{ib})
\Big]\,.
\end{align}
Besides this source, at fourth order in $Y$, there are also the
standard contributions from direct $CP$-violation in the decays and
inverse decays of right-handed neutrinos as well as from their
mixing~\cite{Covi:1996wh}.
These add linearly to the asymmetry from lepton-doublet mixing, that is the
main topic of the present work. For simplicity, we do not account for the
standard sources of asymmetry in most of the following discussion and
refer to Refs.~\cite{Beneke:2010wd,Beneke:2010dz,Garbrecht:2010sz,Garbrecht:2011aw,Drewes:2012ma} for the calculation of these contributions within
the CTP formalism. In Section~\ref{sec:example:strongwashout} however,
we present a comparison of the asymmetry resulting from lepton mixing
to the asymmetry that results from the standard sources.

When expressing the kinetic equation for the lepton doublets as
\begin{align}
\partial_\eta\int\frac{d^4 k}{(2\pi)^4}{\rm i}{\rm tr}[\gamma^0 S^{<,>}_{N ij}(k)]
=-\int\frac{d^4 k}{(2\pi)^4}\frac12{\rm tr}
\left[
{\cal C}_\ell(k)
+{\cal C}_\ell(k)
\right]_{ab}
\end{align}
and observing that
\begin{align}
{\rm tr}[{\cal S}_N]=-g_w{\rm tr}[{\cal S}_\ell]\,,
\end{align}
we see that the sum of the helicity asymmetry within the right-handed
neutrinos and the charge asymmetry in the lepton-doublets is conserved,
unless taking proper
account of the helicity-flipping Majorana masses $M_N$. This
is expected, since $\slashed \Sigma_\ell$, the self-energy
that leads to the mixing of the lepton doublets, does not encompass
lepton-number violating operators in the form of odd powers of the Majorana
masses $M_N$ within the approximation represented by the diagram
in Figure~\ref{figure:collison:source}(c) ({\it i.e.} when accounting
for lepton-doublet mixing only and not for the mixing and the
direct decays of the right-handed neutrinos).

Now, since in the strong washout scenario, $M_{N ii}\gg T$,
the right-handed neutrinos are non-relativistic, and they decay
at tree-level with equal likelihood into leptons and anti-leptons.
This is different for relativistic right-handed neutrinos, that
are converted to leptons via scattering processes with the
particles in the plasma. These reactions are approximately helicity
conserving.
In the following, we calculate the asymmetry in the strong washout scenario,
and for this reason, we only have to account for the production
and the washout of the asymmetry within the leptons $\ell$, because the
helictiy asymmetry within the right-handed neutrinos is rapidly violated by
the Majorana masses.
In the strong washout regime, all energies of particles that participate
in decay and inverse decay processes are much larger than the temperature.
One may therefore approximate Eqs.~(\ref{lepdist:kineq}) by
their Maxwell forms
\begin{subequations}
\label{lepdist:kineq:Maxw}
\begin{align}
\delta f_\ell(\mathbf k)=&\frac{12\delta n_\ell^+}{T^3}{\rm e}^{-|\mathbf k|/T}\,,
\\
\delta \bar f_\ell(\mathbf k)=&\frac{12\delta n_\ell^-}{T^3}{\rm e}^{-|\mathbf k|/T}
\,.
\end{align}
\end{subequations}
We then substitute above approximations into Eqs.~(\ref{sol:deltan}), and these
into the source term for the lepton
asymmetry~(\ref{source:lepto}). It enters the kinetic
equations~(\ref{kin:eq}), when integrating these over $dk^0$,
\begin{align}
\label{source:q}
\partial_\eta q_{\ell aa}
&=\sum\limits_{ic}
\frac{1}{32\pi^3}
\left[
Y^\dagger_{ai}Y_{ic}\frac{6q_{\ell ca}}{T^3}
+\frac{6q_{\ell ac}}{T^3}Y^\dagger_{ci}Y_{ib}
\right]M_{Ni}^3 T K_1(M_{Ni}/T)
+\int\frac{d^4k}{(2\pi)^4}C_{\ell aa}^Y(k)
\\\notag
&=
\sum_{\underset{c\not=a}{ijc}}
\frac{3}{16\pi^3}\frac{M_{Ni}^3}{T^4}K_1(M_{Ni}/T)
B_j^Y {\cal}{\cal Q}_{\ell ca}
{\rm i}
\left(
Y^\dagger_{ai}Y_{ic}Y^\dagger_{cj}Y_{ja}
-Y^\dagger_{aj}Y_{jc}Y^\dagger_{ci}Y_{ia}
\right)
\\
\notag
&
+\int\frac{d^4k}{(2\pi)^4}C_{\ell aa}^Y(k)
\,.
\end{align}
Here, we have evaluated the source term~(\ref{source:lepto}) explicitly
in the limit of strong washout, while the remaining integral corresponds
to the washout term.

\section{Freeze-Out Asymmetries in the Strong Washout Regime}
\label{sec:example:strongwashout}

We now calculate the freeze-out value of the asymmetry produced in the out-of-equilibrium decays of the
individual $N_i$ in the strong washout regime.
The total asymmetry is then obtained as the sum of the
particular asymmetries.
As usual, it is convenient to parametrise the evolution by
\begin{align}
z_i=M_{Ni}/T\,.
\end{align}
The scale factor $a$ can be expressed through the
comoving temperature $a_{\rm R}$ as
\begin{align}
a=a_{\rm R}\eta=a_{\rm R}\frac{z_i}{M_{Ni}}\,,\quad
a_{\rm R}=\frac{m_{\rm Pl}}{2}\sqrt{\frac{45}{\pi^3 g_\star}}\,,
\end{align}
where $g_\star=106.75$ is the number of relativistic degrees of
freedom of the Standard Model at high temperatures.
The physical temperature is given by $T=a_{\rm R}/a=M_{Ni}/z_i$.
In order to take account of the expansion of the
Universe, in all kinetic equations written down to this end,
the mass terms should be multiplied by the scale factor $a$ and
all temperatures be replaced by the comoving temperature $a_{\rm R}$,
{\it cf.} Ref~\cite{Beneke:2010wd}.
The kinetic equations then take the form
\begin{subequations}
\label{kineq:strongwashout}
\begin{align}
\frac{dY^{Ni}_{\ell a}}{dz_i}&=\bar S_{\ell aa}^{Ni} (Y_{Ni}-Y_{Ni}^{\rm eq})
+\bar W_{\ell a} Y_{\ell a}
\,,
\\
\frac{dY_{Nk}}{dz_i}&=\bar{\cal C}_{Nk}(Y_{Nk}-Y_{Nk}^{\rm eq})\,,
\end{align}
\end{subequations}
where $Y_{\ell a}=q_{\ell aa}/s$ and $Y_{Ni}=n_{Ni}/s$, and
$s$ is the entropy density.
The various distributions, collision and source terms are
({\it cf.} Ref.~\cite{Garbrecht:2012qv})
\begin{subequations}
\begin{align}
\bar S_{\ell aa}^{Ni}&=
-\sum\limits_{\underset{j\not=i,\,c\not=a}{jc}}
\frac{3 a_{\rm R}z_i^{\frac92}{\rm e}^{-\frac{M_{Nj}}{M_{Ni}}z_i}}{2^{23/2}\pi^{7/2}}
\frac{M_{Nj}^\frac72}{M_{Ni}^\frac92}
\frac{[YY^\dagger]_{ii}}{[YY^\dagger]_{jj}}
{\cal Q}_{\ell ac}
{\rm i}
\left(
Y^\dagger_{ai}Y_{ic}Y^\dagger_{cj}Y_{ja}
-Y^\dagger_{aj}Y_{jc}Y^\dagger_{ci}Y_{ia}
\right)
\,,
\\
Y_{Nk}^{\rm eq}&=
2^{-1/2}\pi^{-3/2}
\left(\frac{M_{Nk}}{M_{Ni}}\right)^{3/2}
z_i^{3/2}a_{\rm R}^3{\rm e}^{-z_i M_{Nk}/M_{Ni}}/s\,,
\\
\bar{\cal C}_{Nk}&=\frac{g_w}{16\pi}\sum\limits_aY_{ka}Y^\dagger_{ak}a_{\rm R}
z_i\frac{M_{Nk}}{M_{Ni}^2}
\,,
\\
\label{def:B:washout}
\bar W^{Ni}_{\ell a}
&=-\sum\limits_k Y^\dagger_{ak}Y_{ka}
\frac{3 z_i^{5/2}}{2^{9/2}\pi^{5/2}}
\left(\frac{M_{Nk}}{M_{Ni}}\right)^{5/2}
\frac{a_{\rm R}}{M_{Ni}}{\rm e}^{-z_i M_{Nk}/M_{Ni}}
=:\sum\limits_k B^{Ni}_{\ell ak}z_i^{5/2}{\rm e}^{-z_i M_{Nk}/M_{Ni}}\,.
\end{align}
\end{subequations}

To the leading order in deviations from equilibrium,
$Y_{Nj}-Y_{Nj}^{\rm eq}$, the formal solution
to Eqs.~(\ref{kineq:strongwashout}) is given
by~\cite{Kolb:1983ni,Buchmuller:2004nz}
\begin{align}
\label{Yell:formal}
Y^{Ni}_{\ell a}(z_i)&=
\int\limits_0^{z_i} dz^\prime S_{\ell aa}^{Ni}
\frac{\frac{d}{dz^\prime}Y^{\rm eq}_{Ni}}{\bar{\cal C}_{Ni}}
{\rm e}^{\int\limits_{z^\prime}^{z_i}dz^{\prime\prime}\bar W^{Ni}_{\ell a}(z^{\prime\prime})}
\\\notag
&=
\int\limits_0^{z_i} dz^\prime a^{Ni}_{a}(z^\prime)
{\rm e}^{-\int\limits_{z^\prime}^{z_i}dz^{\prime\prime}\sum\limits_k B_{\ell ak}{z^{\prime\prime}}^{5/2}\exp(-z^{\prime\prime} M_{Nk}/M_{Ni})}
\,,
\end{align}
where we define
\begin{align}
a^{Ni}_{a}(z_i)=S_{\ell aa}^{Ni}
\frac{\frac{d}{dz^\prime}Y^{\rm eq}_{Ni}}{\bar{\cal C}_{Ni}}
=&
\sum\limits_{\underset{j\not=i,\,c\not=a}{jc}}
\frac{3 a_{\rm R}^3 z_i^5{\rm e}^{-\frac{M_{Ni}+M_{Nj}}{M_{Ni}}z_i}}{2^8\pi^4 s  [Y Y^\dagger]_{jj}}
\left(\frac{M_{Nj}}{M_{Ni}}\right)^\frac72
\\\notag
\times&
{\cal Q}_{\ell ac}
{\rm i}
\left(
Y^\dagger_{ai}Y_{ic}Y^\dagger_{cj}Y_{ja}
-Y^\dagger_{aj}Y_{jc}Y^\dagger_{ci}Y_{ia}
\right)\,.
\end{align}

For simplicity, we now consider Leptogenesis from two right-handed neutrinos
$N_{i,j}$. This is relevant in the situation when there are only two right-handed neutrinos in the theory,
when additional right-handed neutrinos are much heavier than
$N_1$ and $N_2$ or when these decouple due to the smallness
of their Yukawa couplings.

The exponent in Eq.~(\ref{Yell:formal})
is then extremal for $z^\prime=z_{\rm f}$ with
\begin{align}
\sum\limits_k
B_{\ell ak} z_{\rm f}^{5/2} {\rm e}^{-z_{\rm f} M_{Nk}/M_{Ni}}=&\frac{M_{Ni}+M_{Nj}}{M_{Ni}}\,,
\end{align}
which we solve numerically for $z_{\rm f}$, what determines the freeze-out
temperature.
Laplace's steepest descent method then yields the flavoured
freeze-out asymmetries ({\it cf.} Refs.~\cite{Kolb:1983ni,Buchmuller:2004nz})
\begin{align}
Y^{Ni}_{\ell a}(z=\infty)
=a_{1a}(z_{\rm f})\sqrt{\frac{2\pi}{\sum_k B_{\ell a k}z_{\rm f}^{5/2}{\rm e}^{-z_{\rm f}M_{Nk}/M_{Ni}}}}
{\rm e}^{-\int\limits_{z_{\rm f}}^\infty \sum_k dz^{\prime\prime}B_{\ell a k}{z^{\prime\prime}}^{5/2}\exp(-z^{\prime\prime} M_{Nk}/M_{Ni})}\,.
\end{align}
Finally, the total lepton asymmetry is the sum
\begin{align}
Y_\ell=\sum_{i,a}Y_{\ell a}^{Ni}\,.
\end{align}

In order to assess the relative importance of the asymmetry
from mixing lepton doublets, we compare it to the
lepton asymmetry that results from the wave function and vertex corrections
to the decays of right-handed
neutrinos $N$. We can simply express
\begin{align}
\frac{dY^{Ni}_{\ell a}}{dz}&=\varepsilon^{Ni}_a \bar{\cal C}_{Ni}(Y_{Ni}-Y_{Ni}^{\rm eq})
+\bar W^{Ni}_{\ell a} Y^{Ni}_{\ell a}\,,
\end{align}
where again, $Y_\ell^{Ni}$ is the asymmetry that results
form the decay of $N_i$.
The decay asymmetry is~\cite{Covi:1996wh,Antusch:2011nz}
\begin{align}
\varepsilon^{Ni}_a=
\frac{3}{16\pi [Y Y^\dagger]_{ii}}\sum\limits_{\underset{j\not=i}{j,b}}
\Big\{
{\rm Im}\left[Y^\dagger_{ai}Y^*_{ib}Y^t_{bj}Y_{ja}\right]\frac{\xi(x_j)}{\sqrt{x_j}}
+{\rm Im}\left[Y^\dagger_{ai}Y_{ib}Y^\dagger_{bj}Y_{ja}\right]\frac{2}{3(x_j-1)}
\Big\}\,,
\end{align}
where $x_j=M_{Nj}/M_{Ni}$ and
\begin{align}
\xi(x)=\frac23 x\left[(1+x)\log\frac{1+x}{x}-\frac{2-x}{1-x}\right]
\,.
\end{align}

The freeze-out temperature is now determined by
\begin{align}
\sum\limits_k
B^{Ni}_{\ell ak} z_{\rm f}^{5/2} {\rm e}^{-z_{\rm f} M_{Nk}/M_{Ni}}=&1\,,
\end{align}
where $B^{Ni}_{\ell ak}$ is given by Eq.~(\ref{def:B:washout}).
The factor that appears in the formal solution~(\ref{Yell:formal})
for $Y_{\ell a}^{Ni}$ is now
\begin{align}
a^{Ni}_{a}(z_i)=-\varepsilon^{Ni}_a\frac{a_{\rm R}^3}{s}2^{-\frac12}\pi^{-\frac32}z_i^{\frac32}{\rm e}^{-z_i}\,.
\end{align}

We now pick some particular, but illustrative points in parameter space
in order to compare the asymmetry from the
mixing of lepton doublets to the asymmetry
from the wave function and vertex corrections to the decay
parameter of right-handed neutrinos.
Parameters consistent with neutrino oscillations can be constructed with
the help of the relation~\cite{Casas:2001sr}
\begin{equation}
\label{CasasIbarraDef}
Y^\dagger={V^\Delta}^\dagger U_\nu \sqrt{m_\nu^{\rm diag}} \mathcal{R} \sqrt{M_N} \frac{\sqrt 2}{v},
\end{equation}
where $m_{\nu}^{\rm diag}={\rm diag}(m_1,m_2,m_3)$ is the diagonal mass matrix
of active leptons,
$M_N={\rm diag}(M_{N11},M_{N22},M_{N33})$
and $v=246\,{\rm GeV}$.
The PMNS matrix $U_{\nu}$ is
\begin{equation}
U_\nu=V^{(23)}U_{\delta}V^{(13)}U_{-\delta}V^{(12)}{\rm diag}(e^{i\alpha_1 /2},e^{i\alpha_2 /2},1)
\end{equation}
with $U_{\pm\delta}={\rm diag}(e^{\mp i\delta/2},1,e^{\pm i\delta/2})$. 
The non-zero entries of the matrices $V^{(ij)}$ are
\begin{eqnarray}
V^{(ij)}_{ii}=V^{(ij)}_{jj}=\cos\theta_{ij}\,,\quad
V^{(ij)}_{ij}=\sin\theta_{ij}\,,\quad
V^{(ij)}_{ji}=-\sin\theta_{ij}\,,\quad
V^{(ij)}_ {kk}\underset{k\not=i,j}{=}1\,.
\end{eqnarray}
The $\theta_{ij}$ are the mixing angles
of the active neutrinos, and $\alpha_1$, $\alpha_2$ and $\delta$ are $CP$-violating phases.
The matrix ${\cal R}$ must satisfy
${\cal R}^t{\cal R}=\mathbbm{1}$.
We parametrise it by the complex
angles $\omega_{ij}$ and the matrices ${\cal R}^{(ij)}$
with non-zero entries
\begin{align}
{\cal R}^{(ij)}_{ii}&={\cal R}^{(ij)}_{jj}=\cos\omega_{ij}\,,\quad
{\cal R}^{(ij)}_{ij}=\sin\omega_{ij}\,,\quad
{\cal R}^{(ij)}_{ji}=-\sin\omega_{ij}\,,\quad
{\cal R}^{(ij)}_{kk}\underset{k\not=i,l}{=}1\,,
\\\notag
{\cal R}&={\cal R}^{(23)}{\cal R}^{(13)}{\cal R}^{(12)}\,.
\end{align}
We take $M_{N3}\gg M_{N1,2}$ or alternatively, make the assumption that there
is no right-handed neutrino $N_3$ or simply that the Yukawa couplings
of $N_3$ are negligibly small. The parameter space can then
be restricted by the choice $\omega_{13}=\pi/2$ and $\omega_{23}=0$.
Additional parameters that we choose are listed in Table~\ref{table:params}.
Finally, $V^\Delta$ is a unitary matrix that brings $Y$ to a
form, such that only asymmetries in two flavours $\tau$ and $\sigma$ are produced,
where $\sigma$ is a linear combination of $e$ and $\mu$, see Ref.~\cite{Beneke:2010dz}
for the explicit construction. (Explicitly, this means that
$Y_{2\sigma_\perp}=Y_{3\sigma_\perp}=Y_{3\sigma}=Y_{3\tau}=0$, while the remaining
entries are generically non-zero, and $\sigma_\perp$ is the linear flavour 
combination orthogonal to $\tau$ and $\sigma$.)

\begin{figure}[t!]
\begin{center}
\epsfig{file=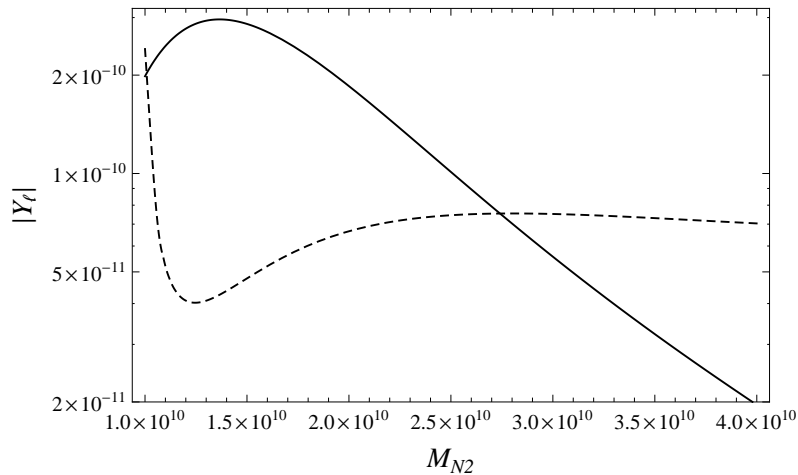,width=10.5cm}
\end{center}
\caption{
\label{fig:leplep}
Lepton asymmetry from mixing leptons (solid line) and from wave-function and vertex
corrections for the right-handed
neutrinos (dashed line). The model parameters are listed in Table~\ref{table:params}.
}
\end{figure}

\begin{table}[t!]
\begin{center}
\begin{tabular}{|r|r|}
\hline
$m_1$ & $0$\\
$m_2$ & $8.7\,{\rm meV}$\\
$\sin\theta_{12}$ & $0.55$\\
$\sin\theta_{23}$ & $0.63$\\
$\sin\theta_{13}$ & $0.16$\\
$\delta$ & $0$\\
\hline
\end{tabular}
\begin{tabular}{|r|r|}
\hline
$\alpha_1$ & $0$\\
$\alpha_2$ & $\pi/2$\\
$\omega_{12}$ & $-\pi/2+{\rm i}/2$\\
$\omega_{13}$ & $\pi/2$\\
$\omega_{23}$ & $0$\\
$M_{N1}$ & $10^{10}\,{\rm GeV}$\\
\hline
\end{tabular}
\end{center}
\caption{\label{table:params}
Parameters for the numerical example in Figure~\ref{fig:leplep}.}
\end{table}

The one remaining parameter is $M_{N2}$, that we vary for values larger than
$M_{N1}$. In Figure~\ref{fig:leplep}, we compare the resulting asymmetry
from
lepton-doublet mixing to the asymmetry from mixing and direct decays of
right-handed neutrinos, where we take account of the decays of
both, $N_1$ and $N_2$.
For $M_{N2}\to M_{N1}$ (degenerate regime), the asymmetry from
the mixing of right-handed neutrinos is strongly enhanced, which is known
as resonant Leptogenesis. The asymmetry from lepton-doublet mixing also
is strongest for $M_{N2}\sim M_{N1}$.
For larger values of $M_{N2}$, we clearly see the exponential suppression of the 
asymmetry, because the kinematic cuts that give rise to the $CP$-violation are
purely thermal and are Maxwell-suppressed for large values of $M_{N2}$ (hierarchical
regime),
{\it cf.} Eq.~(\ref{B:Y}). For intermediate values of $M_{N2}$, between the degenerate
and the hierarchical regimes, the asymmetry from mixing lepton doublets dominates
over the usually considered asymmetry from mixing and direct decays of right-handed
neutrinos.

This example should illustrate that for $M_{N1}\sim M_{N2}$
(more precisely, when the right-handed neutrino masses are within a factor
of a few), there generically is a sizeable contribution to the lepton asymmetry, that originates from lepton mixing. While we have considered here a particular slice of
the parameter space, it would be interesting to perform a more systematic parametric
study of Leptogenesis from mixing lepton doublets in 
the future.

\section{Maximal Resonant Enhancement in Leptogenesis from Multiple Higgs Doublets}
\label{sec:Higgs:mix}

In Ref.~\cite{Garbrecht:2012qv}, Leptogenesis from mixing Higgs doublets is
introduced. The model is based on the Lagrangian
\begin{align}
{\cal L}=&\frac12 \bar \psi_N({\rm i}\slashed\partial - M_N)\psi_N
+\bar \psi_\ell{\rm i}\slashed\partial\psi_\ell
+\sum\limits_k(\partial_\mu \phi_k)(\partial^\mu \phi_k)
\notag\\
-&
\sum\limits_{kl} M^2_{\phi kl}\phi_k^*\phi_l
-\sum\limits_{mk}(Y_{mk}\bar \psi_N \phi_k P_{\rm L}\psi_{\ell m}+{\rm h.c.})
\,.
\end{align}
Notice that the Yukawa couplings $Y$ here are different from those in the
Lagrangian~(\ref{Lagrangian}). Using the same notations and conventions as
in Ref.~\cite{Garbrecht:2012qv}, we obtain the off-diagonal correlations
in the number densities of Higgs and anti-Higgs particles as
\begin{align}
\delta n_{\phi12}^\pm=2\int\limits_0^{\infty,-\infty}
\frac{dk^0}{2\pi}\int\frac{d^3 k}{2\pi} k^0 {\rm i}D_{\phi12}(k)\,,
\end{align}
where $D_\phi$ is the Wightman function for the Higgs field.

The kinetic equations for the off-diagonal correlations of the Higgs particles
are then (we choose a basis where $M_\phi$, that may include
thermal corrections, is diagonal)
\begin{align}
\label{kineq:Higgs}
\pm{\rm i}(M_{\phi11}^2-M_{\phi22}^2)
\delta n_{\phi12}^\pm
=-[Y^\dagger Y]_{12}B_\phi^Y
-B_\phi^{{\slashed{\rm fl}}}\sum_{j}(y_{j1}^2+y_{j2}^2)\delta n_{\phi12}^\pm
-B^g_\phi\left(\delta n_{\phi12}^+ +\delta n_{\phi12}^-\right)\,.
\end{align}
We have assumed here that the flavour-sensitive interactions are dominated by
couplings $y_{ik}$ of the Higgs-fields $\phi_k$ to Standard Model fermions.
Note however that the couplings
$Y$ and the quartic interactions among the Higgs fields are flavour sensitive
as well and should be taken into account when the $y_{ik}$ are not dominating.
An estimate for the weighted annihilation rate $B^g_\phi\approx1.0\times 10^{-4} T^2$
is given
by Eq.~(\ref{Bgphi:int}) in Appendix~\ref{app:rates}. As for
$B^{\slashed{\rm fl}}_\phi$, we
do not provide an estimate here, since it is model dependent. If one of
the Higgs fields $\phi$ is the Standard Model Higgs field, this rate will be dominated
by the top-quark Yukawa couplings $y_{ti}$. Because the scattering processes then
also involve gluons, we may expect this rate to be enhanced by a factor of a
few when compared to $B^{\slashed{\rm fl}}_\ell$.

Eq.~(\ref{kineq:Higgs}) has the solution
\begin{subequations}
\begin{align}
q_{\phi 12}=&\delta n_{\phi12}^+-\delta n_{\phi12}^-
={\cal R}_{\phi}2{\rm i}[Y^\dagger Y]_{12}B_\phi^Y\,,
\\
{\cal R}_{\phi}=&
\frac{M_{\phi11}^2-M_{\phi22}^2}
{
\left(M_{\phi11}^2-M_{\phi22}^2\right)^2
+B_\phi^{{\slashed{\rm fl}}}\sum_{j}(y_{j1}^2+y_{j2}^2)
\left[
B_\phi^{{\slashed{\rm fl}}}\sum_{j}(y_{j1}^2+y_{j2}^2)
+2 B_\phi^g
\right]
}
\,,
\\
B_\phi^Y=&\frac{\mu_N M_N^{7/2} T^{1/2}}{32\sqrt 2 \pi^{5/2}}
{\rm e}^{-M_N/T}\,.
\end{align}
\end{subequations}
The enhancement factor ${\cal R}_{\phi}$ is analogous to the
quantity~(\ref{Qlab})
 ${\cal Q}_{\ell ab}$ for mixing leptons, but we have choose a different
dimensionality for convenience and the sake of comparison with
Ref.~\cite{Garbrecht:2012qv}.

Following Ref.~\cite{Garbrecht:2012qv}, we then obtain for the freeze-out
asymmetry
\begin{subequations}
\begin{align}
\label{Yell:freezeout}
Y_{\ell i}(z=\infty)&=
\frac{{\rm Im}[Y_{i1}Y^*_{j1}Y_{j2}Y^*_{i2}]}{{\rm tr[YY^\dagger]}}
\frac{135\,M_N^2}{64\pi^6 g_\star}{\cal R}_\phi
\sqrt{\frac{{\rm e}^{z_{{\rm f}i}}}{\sqrt{z_{{\rm f}i}}B_i}}
{\rm e}^{-2z_{{\rm f}i}-\int\limits_{z_{{\rm f}i}}^\infty dz B_i z^{5/2}{\rm e}^{-z}}\,,
\,\\
B_i&=[Y^\dagger Y]_{ii}
\frac{9}{32\pi^4}\sqrt{\frac{5}{2g_\star}}\frac{m_{\rm Pl}}{M_N}\,,
\,\\
z_{{\rm f}i}&=-\frac{5}{2}W_{-1}\left((-{2}/{5})\times\left({2}/{B_i}\right)^{2/5}\right)\,.
\end{align}
\end{subequations}
(Note a missing factor $1/\sqrt{B_i}$ in the expression for $Y_{\ell i}$
in Ref.~\cite{Garbrecht:2012qv}.)

\section{Summary and Discussion}
\label{sec:sum:disc}

We have calculated the mixing of Standard Model lepton doublets in the early
Universe and studied its impact on Baryogenesis via Leptogenesis. The mixing
is driven by non-equilibrium right-handed neutrinos, as described
in Section~\ref{sec:off:corr}. With no deviation
from equilibrium, there is no mixing of the lepton doublets. The amount of 
mixing is quantified by the the correlations of the off-diagonal number
densities~(\ref{sol:deltan}).

Substituting these off-diagonal flavour correlations back into the
collision term for the leptons, we obtain the source for the
lepton asymmetry, Eqs.~(\ref{source:lepto},\ref{source:q}) in
Section~\ref{sec:cancellation}. This asymmetry
is opposite to the source for the helicity asymmetry in the right-handed
neutrinos (which is however rapidly violated through the Majorana masses).

By calculating the freeze-out value of the lepton asymmetry, we compare
the source from lepton mixing to the standard source from mixing and
direct decays
of right-handed neutrinos in Section~\ref{sec:example:strongwashout}.
The standard result is enhanced for mass-degenerate right-handed neutrinos
(resonant Leptogenesis), while the asymmetry from lepton mixing is
Maxwell-suppressed for larger mass ratios. In between these regimes,
we find that lepton-doublet mixing may be the dominant source of
$CP$-asymmetry for Leptogenesis.

Results for the impact of gauge interactions on the mixing of Higgs-doublets
and implications for the freeze-out value of the induced asymmetries are
presented in Section~\ref{sec:Higgs:mix}.

In the future, it would be interesting to perform a more systematic scan
of parameter space, as the asymmetries presented in Figure~\ref{fig:leplep}
are from a particular slice only. For that purpose,
a more systematic and accurate calculation of the rates of right-handed
neutrino production and pair creation and annihilation of doublet
leptons, that we estimate here in Appendix~\ref{app:rates}, would be beneficial.
The relevant techniques are currently under development, see
Refs.~\cite{Anisimov:2010gy,Besak:2012qm,Salvio:2011sf,Laine:2011xm,Laine:2011pq}.

The Closed-Time-Path method so far has already proven useful
in order to describe the emergence of the baryon asymmetry of the Universe 
in a more systematic manner and
in order to calculate certain finite-temperature
corrections to the approximations that are most commonly
applied~\cite{Garbrecht:2011aw,Garny:2011hg,Drewes:2012ma,Garbrecht:2010sz,Buchmuller:2000nd,De Simone:2007rw,Garny:2009rv,Garny:2009qn,Anisimov:2010aq,Garny:2010nj,Beneke:2010wd,Beneke:2010dz,Garny:2010nz,Anisimov:2010dk}
.
Together with Ref.~\cite{Garbrecht:2012qv}, the present work shows that
these methods can also be useful in order to identify sources
for the $CP$-asymmetry, that have not been considered before. The further
development of non-equilibrium field theory techniques and their
application to the comparably simple problem of Leptogenesis will therefore
prove helpful for progress on both, the qualitative and quantitative
understanding of the emergence of the baryon-asymmetric Universe.

\subsection*{Acknowledgements}
The author acknowledges
support by the Gottfried Wilhelm Leibniz programme
of the Deutsche Forschungsgemeinschaft.

\begin{appendix}
\section{Pair Creation and Annihilation Rates and Flavour Sensitive Rates}
\label{app:rates}

Pair creation and annihilation can be mediated either through the $s$-channel or the 
$t$-channel. Individual $s$-channel processes are typically suppressed compared to
$t$-channel processes by a factor of a few (essentially the logarithm of the
squared gauge coupling), because the latter may be infrared enhanced due to the
exchange of soft particles. Nonetheless, due to the large number of
matter degrees of freedom in the Standard Model, it should be a sufficient approximation to consider
$s$-channel processes for pair creation and annihilation of matter
particles and Higgs bosons only. Technically,
it should also be possible to account for the $t$-channel contributions, which would 
however be more complicated due to the infrared enhancement and its screening.

A standard procedure for calculating cosmological
interaction rates of non-relativistic particles
relies on the reduced cross section $\hat \sigma(s)$, where $s$ is the Mandelstam
$s$-variable. The main simplification arises due to the use of Maxwell statistics.
For massless particles, this approximation incurs an order one inaccuracy. Moreover,
the $t$-channel contributions may leave the reduced cross section infrared divergent,
a problem that should be accounted for by including the screening effects in the 
plasma. In conjunction with the comments above, we should expect that the following
procedure yields an ${\cal O}(1)$ estimate of the weighted pair creation and
annihilation rate $B_\ell^g$. A more accurate calculation of this quantity may be
subject of a future study.

The total annihilation rate of massless particles $A$ into massless
matter or Higgs particles $B$
in Maxwell approximation is given by
\begin{align}
\gamma_{A\bar A\to B\bar B}
&=
\int\frac{d^3 k_A}{2|\mathbf k_A|}
\frac{d^3 k_{\bar A}}{2|\mathbf k_{\bar A}|}
\frac{d^3 k_B}{2|\mathbf k_B|}
\frac{d^3 k_{\bar B}}{2|\mathbf k_{\bar B}|}
{\rm e}^{-|\mathbf k_A|/T}{\rm e}^{-|\mathbf k_{\bar B}|/T}
|{\cal M}_{A\bar A\to B\bar B}|^2
\\\notag
&=\frac{T}{64\pi^4}\int\limits_0^\infty ds \sqrt s K_1\left(\frac{\sqrt s}{T}\right)
\hat\sigma(s)
\,.
\end{align}
The squared matrix element is understood to be a sum over all external polarisation
and weak isospin states. The reduced cross section is defined as
\begin{align}
\hat\sigma(s)=\frac{1}{8\pi s}\int\limits_0^\infty dt |{\cal M}_{A\bar A\to B\bar B}|^2\,.
\end{align}

Let $f_{A,B}$ denote chiral fermions, $b_{A,B}$ bosons. We obtain that
\begin{subequations}
\begin{align}
\gamma_{f_A\bar f_A\to f_B\bar f_B}&=G \frac{T^4}{24\pi^5}\,,\\
\gamma_{f_A\bar f_A\to b_B\bar b_B}&=G \frac{T^4}{192\pi^5}\,,\\
\gamma_{b_A\bar b_A\to b_B\bar b_B}&=G \frac{T^4}{384\pi^5}\,.
\end{align}
\end{subequations}
When $f_{X,Y}$ and $b_{X,Y}$ are ${\rm SU}(2)_{\rm L}$ doublets,
$G=\frac34 g_2^4 + g_1^4(q_A^2 q_B^2)$, where $q_{A,B}$ denotes the ${\rm U}(1)_Y$
weak hypercharge of the particles. For singlets,
$G=g_1^4(q_A^2 q_B^2)$.
For the doublets, we have to divide the rates by two in order to average over
the weak isospin such that we obtain the production rate for a single isospin state.
Summing over all Standard Model degrees of freedom for $B$, we obtain

\begin{align}
\gamma_{\ell\bar\ell\to\textnormal{anything}}
=\left(
\frac{97}{512\pi^5}g_2^4
+\frac{163}{4608 \pi^5} g_1^4
\right)T^4\,.
\end{align}
The averaged rate for a single flavour-sensitive 
process is therefore~\cite{Drewes:2012ma}
\begin{align}
\Gamma^{g}_{\rm av}=
\frac{\gamma_{\ell\bar\ell\to\textnormal{anything}}}{3/(2\pi)^2 \zeta(3)T^3}
\approx 4.6\times 10^{-4}T\,.
\end{align}
We have taken here for the numerical values $g_2=0.6$ and $g_1=0.4$ at
the time of Leptogenesis.
Now with the order one
approximation that all lepton doublets experience the same reaction rate,
regardless of their momentum, we express
\begin{align}
\label{Bg:int}
B_\ell^{g} n^+_\ell
=\int\frac{d^3 k}{(2\pi)^3}|\mathbf k|\Gamma^{\slashed{\rm fl}}_{\rm av} f_{\ell}
=\delta n^+_\ell\frac{12\Gamma^{\slashed{\rm fl}}_{\rm av}}{T^3}
\int\frac{d^3 k}{(2\pi)^3}|\mathbf k|{\rm e}^{-|\mathbf k|/T}\,,
\end{align}
such that it follows that
\begin{align}
B_\ell^{g}=\frac{36}{\pi^2}\Gamma^{g}_{\rm av}T=1.7\times10^{-3} T^2.
\end{align}
The difference when using a Fermi-Dirac instead of a Maxwell distribution
in the above integral~(\ref{Bfl:int}) is, due to the factor $|\mathbf k|$ that
favours contributions with large momentum, only about 6\%.
Using the same methods, we obtain for the weighted annihilation rate of Higgs
particles
\begin{align}
\label{Bgphi:int}
B^g_\phi=1.0\times 10^{-4} T^2\,.
\end{align}

As for the flavour sensitive rates, we proceed similarly and
make our present estimates as follows:
Ref.~\cite{Besak:2012qm} provides the number for the total production rate
for sterile right-handed neutrinos. This gives an approximation for the flavour-sensitive scatterings that involve charged right-handed
charged leptons. The main
difference is that the charged right-handed leptons may also radiate ${\rm U}(1)_Y$
gauge bosons, an effect that is therefore neglected when adopting the number
$h^\dagger h 5\times 10^{-4} T^4$ for the total reaction rate
from Ref.~\cite{Besak:2012qm}. The averaged rate for a single flavour-sensitive 
process then is
\begin{align}
\Gamma^{\slashed{\rm fl}}_{\rm av}=
\frac{5\times10^{-4}}{3/(2\pi)^2 \zeta(3)}T\approx 3\times 10^{-3}T\,.
\end{align}
In the approximation of the same reaction rate for all leptons
in phase space, we obtain
\begin{align}
\label{Bfl:int}
B_\ell^{\slashed{\rm fl}} h^\dagger h \delta n^+_\ell
=\int\frac{d^3 k}{(2\pi)^3}|\mathbf k|\Gamma^{\slashed{\rm fl}}_{\rm av} h^\dagger h\delta f_{\ell}
=h^\dagger h\delta n^+_\ell\frac{12\Gamma^{\slashed{\rm fl}}_{\rm av}}{T^3}
\int\frac{d^3 k}{(2\pi)^3}|\mathbf k|{\rm e}^{-|\mathbf k|/T}\,,
\end{align}
and therefore
\begin{align}
B_\ell^{\slashed{\rm fl}}=\frac{36}{\pi^2}\Gamma^{\slashed{\rm fl}}_{\rm av}T=1.0\times10^{-2} T^2.
\end{align}

In case a more accurate calculation of the lepton asymmetries is desired
in the future, the estimates for the rates made in this Appendix, which
should be of ${\cal O}(1)$ accuracy, need to be improved. The main challenges
are then to account for infrared effects and to abandon the approximations
of averaging the reaction rates over phase space.

\end{appendix}

\end{document}